\documentclass[sigconf]{acmart}
\usepackage{graphicx}
\usepackage{amsmath}
\usepackage[normalem]{ulem}
\usepackage{algorithm}
\usepackage{algpseudocode}
\usepackage{diagbox}
\usepackage{multirow}
\usepackage{subfig}
\usepackage{float}
\usepackage{balance}
\useunder{\uline}{\ul}{}
\AtBeginDocument{%
  }

\setcopyright{acmcopyright}
\copyrightyear{2018}
\acmYear{2018}
\acmDOI{XXXXXXX.XXXXXXX}

\acmConference[Conference acronym 'XX]{Make sure to enter the correct
  conference title from your rights confirmation emai}{June 03--05,
  2018}{Woodstock, NY}
\acmPrice{15.00}
\acmISBN{978-1-4503-XXXX-X/18/06}

\begin{document}

\title{Interaction-level Membership Inference Attack Against Federated Recommender Systems}

\author{Wei Yuan}
\affiliation{%
  \institution{The University of Queensland}
  \city{Brisbane}
  \country{Australia}
}
\email{w.yuan@uq.edu.au}

\author{Chaoqun Yang}
\affiliation{%
  \institution{Griffith University}
  \country{Gold Coast, Australia}}
\email{chaoqun.yang@griffith.edu.au}

\author{Quoc Viet Hung Nguyen}
\affiliation{%
  \institution{Griffith University}
  \city{Gold Coast}
  \country{Australia}
}
\email{henry.nguyen@griffith.edu.au}

\author{Lizhen Cui}
\affiliation{%
 \institution{Shandong University}
 \city{Jinan}
 \country{China}}
 \email{clz@sdu.edu.cn}

\author{Tieke He}
\affiliation{%
  \institution{Nanjing University}
  \city{Nanjing}
  \country{China}}
  \email{hetieke@gmail.com}

\author{Hongzhi Yin}
\authornote{Corresponding author.}
\affiliation{%
  \institution{The University of Queensland}
  \city{Brisbane}
  \country{Australia}}
\email{h.yin1@uq.edu.au}

\renewcommand{\shortauthors}{Yuan et al.}

\begin{abstract}
  The marriage of federated learning and recommender system (FedRec) has been widely used to address the growing data privacy concerns in personalized recommendation services.
  In FedRecs, users' attribute information  and behavior data (i.e., user-item interaction data) are kept locally on their personal devices, therefore, it is considered a fairly secure approach to protect user privacy.
  As a result, the privacy issue of FedRecs is rarely explored.
  Unfortunately, several recent studies reveal that FedRecs are vulnerable to user attribute inference attacks, highlighting the privacy concerns of FedRecs.
  In this paper, we further investigate the privacy problem of user behavior data (i.e., user-item interactions) in FedRecs.
  Specifically, we perform the first systematic study on interaction-level membership inference attacks on FedRecs.
   An interaction-level membership inference attacker is first designed, and then the classical privacy protection mechanism, Local Differential Privacy (LDP), is adopted to defend against the membership inference attack. Unfortunately, the empirical analysis shows that LDP is not effective against such new attacks unless the recommendation performance is largely compromised.
  To mitigate the interaction-level membership attack threats, we design a simple yet effective defense method to significantly reduce the attacker's inference accuracy without losing recommendation performance.
  Extensive experiments are conducted with two widely used FedRecs (Fed-NCF and Fed-LightGCN) on three real-world recommendation datasets (MovieLens-100K, Steam-200K, and Amazon Cell Phone), and the experimental results show the effectiveness of our solutions.
\end{abstract}




\maketitle

\section{Introduction}\label{sec:introduction}
In the age of the information explosion, recommender systems have become an essential means to alleviate information overload~\cite{gomez2015netflix,yin2017spatial}, and many recommendation techniques have been proposed, including matrix factorization~\cite{mnih2007probabilistic}, deep learning based methods~\cite{he2017neural,he2020lightgcn}, etc.
These traditional recommender systems have already achieved good performance in diverse scenarios~\cite{zhang2019deep}.
However, most of these traditional recommender systems work in a centralized way, i.e., they require collecting and storing users' historical interaction data  to train a powerful recommender model in a central server~\cite{lam2006you}.
As the increasing concerns of user privacy and the relevant privacy protection regulations such as the General Data Protection Regulation (GDPR)~\cite{voigt2017eu} in European Union and the California Consumer Privacy Act (CCPA)~\cite{harding2019understanding} in the United States, centrally collecting users' personal data is harder and even becomes infeasible in many cases~\cite{liang2021fedrec++}.

To address the privacy issue, federated learning (FL)~\cite{mcmahan2017communication} has been recently adopted in recommender systems.
In federated recommender systems (FedRecs), users can collaboratively train the recommender model but do not need to share their private data with either central servers or other users (clients).
Therefore, FedRecs are  considered a natural solution to protect users' sensitive information.
Generally, FedRecs can be further divided into FedRecs with explicit feedback~\cite{liang2021fedrec++} and FedRecs with implicit feedback, according to their training datasets and optimization objectives.
In this paper, we focus on FedRecs with implicit feedback\footnote{To make the presentation concise, we directly use FedRec refer to FedRec with implicit feedback by default in the remaining part of this paper.}.
Since Ammad et al.~\cite{ammad2019federated} proposed the first FedRec with collaborative filtering, many studies followed and extended their basic FedRec framework.
For example, FedFast~\cite{muhammad2020fedfast} aims to accelerate the convergence of FedRec training.
Imran et al.~\cite{imran2022refrs} and Wang et al.~\cite{wang2022fast} focused on the efficiency of FedRecs.

With the remarkable attainment achieved in a short time~\cite{yang2020federated}, a few recent studies have started to verify whether FedRecs are ``safe'' enough.
\cite{zhang2022comprehensive} is the first work to analyze the privacy issue of FedRecs.
However, it only discussed sensitive attribute information leakage~\cite{zhang2021graph} and developed an effective attribute information protection approach.
Although \cite{lin2020fedrec,liang2021fedrec++,lin2021fr,lin2022generic} studied the leakage and protection of user rating information in FedRecs, they all focused on explicit feedback data,  which are much different from this work targeting FedRecs with implicit feedback. 

Inferring a user's interaction data in FedRecs is one type of membership inference attack (MIA).  Although MIA has been widely investigated in federated classification tasks~\cite{nasr2019comprehensive,zhang2020gan,geiping2020inverting,lyu2020threats,zhao2021user,suri2022subject}, their proposed attack and defense approaches cannot apply to  FedRecs due to the following major differences between federated recommendation  and federated classification.  (1) From the perspective of attack objective,  MIA in federated classification aims to infer or predict whether a sample has been used in the federated training process and which client has used it for the local training.
However, in FedRecs, the associated item set of each client can be easily inferred  by simply checking which items' embeddings are updated by the client.  Furthermore, knowing such an item set is meaningless in FedRecs, since it consists of both positive and negative samples/items, and only positive samples (i.e., interacted items) can leak user privacy.
Hence, the membership inference attack on FedRecs aims to infer the user's interacted items (i.e., positive samples), and we name such MIA as Interaction-level Membership Inference Attack (IMIA).
(2) From the attack implementation perspective, MIA in federated classification needs to acquire extra i.i.d. data, which is however infeasible in FedRecs. In addition, the federated recommender architecture is significantly different from the federated classification model architecture. A client in FedRecs can have its private parameters (i.e., user embedding), while all model parameters in the federated classification models are shared.

In this paper, we first design a novel IMIA attacker to reveal the risk of leaking user interaction data in FedRecs and then  propose an efficient and effective defender. The attack is launched by a central server that is honest but curious.
The central server aims to identify a user's interacted items (i.e., positive samples) from its associated items (including both positive and negative samples) by analyzing the user's uploaded parameters without breaking the federated learning protocol.
To be specific, given a target client, the attacker iteratively identifies its interacted items by repeating the following procedure.  The attacker first randomly assigns ratings (0 or 1) to the client's associated items to construct a shadow training set, based on which a shadow recommender model is trained.  Then, the attacker compares the relevance between the client's uploaded item embeddings and the item embeddings in the shadow recommender model to find the correctly guessed items. We implement the IMIA attacker on 
two representative FedRecs (Fed-NCF~\cite{ammad2019federated} and Fed-LightGCN~\cite{he2020lightgcn}), and evaluate its inference accuracy on 
  three real-world recommendation datasets (MovieLens-100K~\cite{harper2015movielens}, Steam-200K~\cite{cheuque2019recommender}, and Amazon Cell Phone~\cite{he2016ups}). 
The experimental results show the high inference accuracy of this new IMIA attacker, highlighting the risk of user interaction data leakage in FedRecs.

Recently, to improve the privacy-preserving ability of federated learning, Local Differential Privacy (LDP) has been employed in FedRecs and quickly becomes a gold standard for privacy preservation because of its effectiveness~\cite{wang2019collecting,yang2020local,liu2022federated}.
Therefore, we also evaluate the performance of the IMIA attacker in the above-mentioned FedRecs equipped with LDPs.
It is found that LDP is not effective against such new attacks unless the recommendation performance is largely compromised, highlighting the timely demand for a new defense mechanism against the new IMIA.

In light of this, we propose a novel defense mechanism - IMIA defender. As there are both public and private parameters in FedRecs and only the public parameters can leak user privacy information, we impose a regularization term in the loss function of FedRecs to restrict the update and learning ability of the public parameters and enforce the private parameters to learn more useful patterns and account more for the recommendation performance.
In this way, less sensitive information is transmitted to the server via the shared parameters. As shown in our experiments, our proposed defender can significantly decrease the inference accuracy of the IMIA attacker to the level of random guess with negligible influence on the recommendation performance.

In conclusion, the main contributions of this paper are summarized as follows:
\begin{itemize}
  \item To the best of our knowledge, we are the first to perform a comprehensive privacy analysis of federated recommender systems under interaction-level membership inference attack (IMIA). Our study discloses the privacy risk of user interaction data in FedRecs.
  \item We find that the commonly used privacy-preserving approach, LDP, cannot effectively defend against the new IMIA attack. Then, we propose  a simple yet effective defense mechanism to constrain the update of public parameters, which can significantly degenerate the IMIA attacker's performance to the level of random guesses without hurting the recommendation performance.  
  \item Extensive experiments are conducted with two widely used federated recommender systems (Fed-NCF and Fed-LightGCN) on three real-world recommendation datasets, showing the effectiveness of our attack and defense approaches.
\end{itemize}
\section{Preliminaries}
In this section, we first revisit the fundamental settings of FedRecs, and then formally define interaction-level membership inference attack and defense.
Note that the bold lowercase (e.g. $\mathbf{a}$) represents vectors, bold uppercase (e.g. $\mathbf{A}$) denotes matrices, and squiggle uppercase (e.g. $\mathcal{A}$) signifies sets.

\begin{figure}[!htbp]
  \centering
  \includegraphics[width=0.48\textwidth]{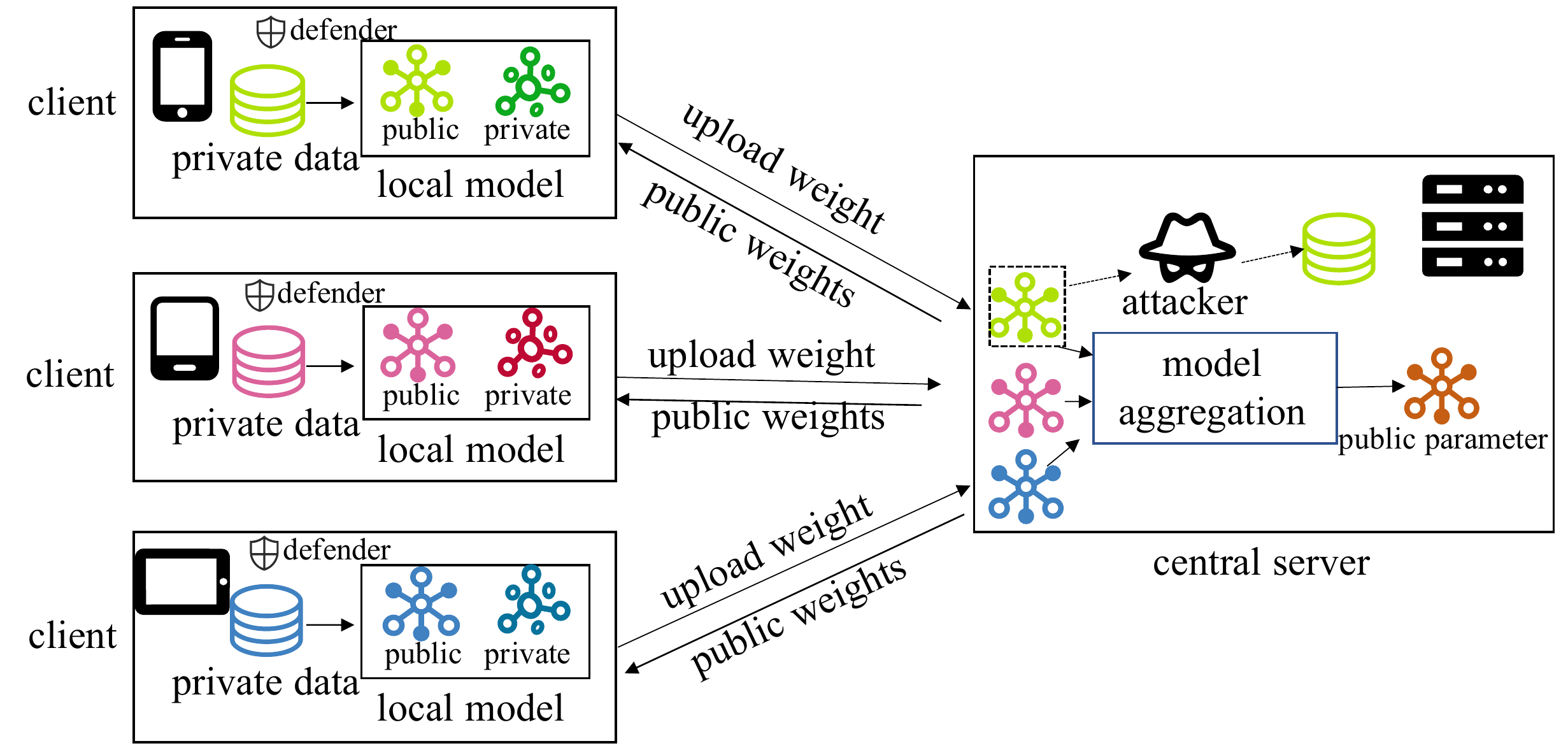}
  \caption{A typical federated recommender system with IMIA attacker and defender.}\label{fig_fed_learning}
\end{figure}

\subsection{Federated Recommender System}
Let $\mathcal{U}$ and $\mathcal{V}$ denote the sets of users (clients) and items, respectively. 
In FedRec, each user/client $u_{i}$ has a local training dataset $\mathcal{D}_{i}$, which consists of user-item interactions $(u_{i}, v_{j}, r_{ij})$.
$r_{ij}=1$ means that user $u_{i}$ has interacted with item $v_{j}$; otherwise, $r_{ij}=0$, that is $v_{j}$ is a negative sample.
We use $\mathcal{V}_{i}^{+}$ and $\mathcal{V}_{i}^{-}$ to denote the interacted item set and negative sample set of user $u_{i}$.
The FedRec is trained to predict $\hat{r}_{ij}$ between $u_{i}$ and non-interacted items.
Finally, FedRec will recommend top-$K$ ranked items with the highest predicted ratings to each user $u_i$.

In FedRec, a central server coordinates a large number of clients. 
The federated training process mainly contains four steps.
First, the central server randomly selects a batch of users/clients as participants and dispenses the global parameters to these clients.
Second, after receiving global parameters, each client combines these public parameters with their private parameters to form a local recommendation model and optimize this model on their local datasets regarding a certain objective function (e.g., BPRLoss~\cite{rendle2012bpr}).
Third, after local training, each client sends the updated public parameters back to the central server.
Finally, the central server aggregates received public parameters with a certain aggregation strategy (e.g., FedAvg~\cite{mcmahan2017communication}).
The above steps form a global training epoch in FedRec and will be repeated many times until the model convergence or meet some pre-defined requirement.

\subsection{Interaction-level Membership Inference Attack and Defense}
\textbf{Adversary's Goal.} 
In this paper, we assume the central server is honest-but-curious, i.e., the server is curious about user private data, but it will not break FedRec's learning protocol.
The goal of the curious server is to infer the set of interacted items on each client $u_{i}$ based on its uploaded public parameters:
\begin{equation}
  \hat{\mathcal{V}}_{i}^{+} \leftarrow IMIA(\mathbf{V}_{i}^{t})
\end{equation}
where $\hat{\mathcal{V}}_{i}^{+}$ is the inferred set of $u_i$'s interacted items, and $\mathbf{V}_{i}^{t}$ represents public or shared parameters that user $u_{i}$ sends to the server at epoch $t$.
Without loss of generality, the public parameters mainly refer to item embeddings in this paper.
The central curious server aims to accurately infer each client's interacted items, and meanwhile, it does not expect its inference attack to affect FedRec's normal learning process and recommendation performance.

\textbf{Adversary's Knowledge.} 
To be more realistic, we assume that the server has the following prior knowledge:  (1) the target user $u_{i}$'s uploaded public parameters (or gradients), which is consistent with the FedRec protocol;  and
(2) a few basic learning hyper-parameters, such as learning rate $lr$ and the ratio of negative sampling $\eta$.  In FedRecs, these hyper-parameters are pre-defined by the central server and broadcast to each participant client, therefore, this assumption of prior knowledge is reasonable.

\textbf{Defense.}
The defense is launched locally by each client to defend against the curious server's inference attack.
The client anticipates the defense method can significantly reduce the server's inference accuracy  to protect their interaction data  
without much recommendation performance loss and  extra computation footprint. 

\section{Method}
In this section, we will first describe the base federated recommenders used in this paper and then present the details of the IMIA attacker and defender. Fig.~\ref{fig_fed_learning} shows the framework of FedRec with IMIA attack and defense and the whole procedure is also described in Alg.~\ref{alg_fedrec}.  

\subsection{Base Federated Recommender}
Generally, a federated learning framework can be applied to most deep learning-based recommendation models.
Among these recommenders, neural collaborative filtering (NCF)~\cite{he2017neural} and graph neural network (GNN)~\cite{scarselli2008graph} are the two most widely used techniques. Hence, we extend an NCF-based centralized model and a LightGCN-based~\cite{he2020lightgcn} centralized model to Fed-NCF and Fed-LightGCN respectively, which will be then used as our base FedRecs to show the effectiveness of our attacker and defender.

\textbf{Neural Collaborative Filtering.} NCF extends collaborative filtering (CF) by leveraging an $L$-layer feedforward network (FFN) to capture the complex patterns of user-item interactions as follows: 
\begin{equation}\label{eq_predict}
  \hat{r}_{ij} = \sigma (\mathbf{h}^\top FFN([\mathbf{u}_{i}, \mathbf{v}_{j}]))
\end{equation}
where $\mathbf{u}_{i}$ and $\mathbf{v}_{j}$ are user $u_{i}$'s and item $v_{j}$'s embedding; $\mathbf{h}$ denotes a learnable weight vector; $[\cdot]$ is concatenation operation, and $\hat{r}_{ij}$ is the predicted preference score of user $u_i$ on item $v_{j}$. 

\textbf{LightGCN.} In graph-based recommenders, the user-item interactions can be constructed as a bipartite graph. Then, LightGCN treats all users and items as distinct nodes.
After that, user and item embeddings are learned by propagating their neighbor nodes' embeddings:
\begin{equation}
  \label{eq_lightgcn}
    \mathbf{u}_{i}^{l} = \sum\limits_{j\in \mathcal{N}_{u_{i}}}\frac{1}{\sqrt{\left| \mathcal{N}_{u_{i}} \right|} \sqrt{\left| \mathcal{N}_{v_{j}} \right|}}\mathbf{v}_{j}^{l-1},\quad
    \mathbf{v}_{j}^{l} = \sum\limits_{i\in \mathcal{N}_{v_{j}}}\frac{1}{\sqrt{\left| \mathcal{N}_{v_{j}} \right|} \sqrt{\left| \mathcal{N}_{u_{i}} \right|}}\mathbf{u}_{i}^{l-1}
\end{equation}
where $\mathcal{N}_{u_{i}}$ and $\mathcal{N}_{v_{j}}$ denote the sets of $u_i$'s and $v_j$'s neighbors.  $l$ is the propagation layer.
Note that under the federated learning setting, each user/client can only access its own data, thus they can only perform the above calculation on their local bipartite graphs.

After $L$ layers propagation, we aggregate all layers' embedding together as the final user and item embeddings:
\begin{equation}
    \mathbf{u}_{i} = \sum\limits_{l=0}^{L} \mathbf{u}_{i}^{l}, \quad \mathbf{v}_{j} = \sum\limits_{l=0}^{L} \mathbf{v}_{j}^{l} 
\end{equation}
Then, as done in NCF,  E.q.~\ref{eq_predict} is adopted to compute the predicted preference scores.

\textbf{FedRec Learning Protocol.}
In FedRec, the parameters can be divided into private and public parameters.
Each client initializes its private parameters, i.e., user embedding $\mathbf{u}_{i}$, and the public parameters $\mathbf{V}$ are initialized by a central server $s$.
At the beginning of a global training epoch $t$, the server $s$ randomly selects a group of clients as participants $\mathcal{U}_{t}$ and sends $\mathbf{V}_{t}$ to each participant. 
The participant combines $\mathbf{V}_{t}$ with its private parameters to form a local recommender and trains the recommender on its local dataset $\mathcal{D}_{i}$ with the following loss function:
\begin{equation}\label{eq_ori_loss}
  \mathcal{L}^{rec} = -\sum\nolimits_{(u_{i}, v_{j}, r_{ij})\in \mathcal{D}_{i}} r_{ij}\log \hat{r}_{ij} + (1-r_{ij})\log (1-\hat{r}_{ij})
\end{equation}
After the local training, the client $u_{i}$ locally updates its private user embedding $\mathbf{u}_{i}$ and uploads the updated public parameters $\mathbf{V}_{i}^{t}$ to the central server $s$.
Then, the server utilizes FedAvg~\cite{mcmahan2017communication} to update the global parameters:
\begin{equation}
  \mathbf{V}_{t+1} = \sum\limits_{u_{i} \in \mathcal{U}_{t}} \mathbf{V}_{i}^{t}
\end{equation}
The above steps iterate until the system converges or meets certain requirements.

\textbf{Local Differential Privacy.} 
As one of the most popular ways to protect users' sensitive data, LDP has been integrated into many FedRecs~\cite{wu2021personalized}.
In this paper, we perform the analysis of IMIA attacks on not only the vanilla FedRecs but also  FedRecs with the LDP mechanism.
Following~\cite{wei2020federated}, before uploading public parameters to server $s$, the client adds some noises to $\mathbf{V}_{i}^{t}$:
\begin{equation}\label{eq_ldp}
  \mathbf{V}_{i}^{t} \leftarrow \mathbf{V}_{i}^{t} + \mathcal{N}(\mathbf{0}, \lambda^{2}\mathbf{I})
\end{equation}
where $\mathcal{N}$ is the normal distribution and $\lambda$ controls the scale of noise.

\subsection{Interaction-level Membership Inference Attacker}\label{sec_imia_attack}
In this work, the curious-but-honest central server is the IMIA attacker, who attempts to infer target user $u_{i}$'s interacted item set $\mathcal{V}_{i}^{+}$.
Basically, if the server has more prior information, such a membership attack is easier to implement with high accuracy.
For example, if the server $s$ can access $u_{i}$'s private user embedding or a part of $u_{i}$'s interaction data, it can simply train a shadow recommender to infer its other interacted items.
However, these strong prior knowledge assumptions are unrealistic in real-world FedRecs.
 Therefore, we assume that the malicious server can only access the public parameters $\mathbf{V}_{i}^{t}$ uploaded by each client  and some training hyper-parameters including the learning rate $lr$ and the negative sampling ratio $\eta$.

Based on the public parameters $\mathbf{V}_{i}^{t}$ updated by $u_i$, the  server can easily infer which items are involved during the local training according to their embedding updates. That is, for item $v_j$, if its embedding is updated by the client $u_i$, $v_j$ participates in $u_i$'s local training. But such simple inference is not useful since $v_{j}$ can also be a negative sample.
The malicious server would like to further infer whether $v_{j}$ is positive or not for user $u_{i}$  (i.e., the value of $r_{ij}$).
Once the $r_{ij}$ is accurately predicted, $u_i$'s private interaction dataset $\mathcal{D}_{i}$ is exposed to the server.
Thus, the membership inference attack problem transforms to predict $r_{ij}$ for item $v_{j}$ in $\mathcal{V}_{i}$.

Our attacker design is inspired  by the following interesting empirical observation.
Assume there is a local model $\mathbf{M}_{i}$ trained on its local dataset $\mathcal{D}_{i}$.
$\mathbf{M}_{i}^{'}$ is also trained on $\mathcal{D}_{i}$ but its private parameters (i.e., user embedding)  have different initial values.
$\mathcal{D}_{i}^{j}$ represents a dataset in which  $v_{j}$'s rating $r_{ij}$ is reversed, and all the other ratings are the same as in $\mathcal{D}_{i}$.
For example, if $r_{ij}=1$ in $\mathcal{D}_{i}$, $r_{ij}$ will be reversed to $0$ in $\mathcal{D}_{i}^{j}$.
$\mathbf{M}_{i}^{''}$ is trained on $\mathcal{D}_{i}^{j}$ with a different private parameter initial point.
Before training, these three models' public parameters are the same.
After training, we obtain the following interesting observation: $dist(\mathbf{v}_{j}, \mathbf{v}_{j}^{'}) < dist(\mathbf{v}_{j}, \mathbf{v}_{j}^{''})$. $dist(\cdot)$ denotes a distance function and the Euclidean metric is adopted in our paper. 
$\mathbf{v}_{j}$, $\mathbf{v}_{j}^{'}$, and $\mathbf{v}_{j}^{''}$ are $v_{j}$'s embeddings from model $\mathbf{M}_{i}$, $\mathbf{M}_{i}^{'}$, and $\mathbf{M}_{i}^{''}$, respectively.
It is worth noting that $u_i$'s embeddings in $\mathbf{M}_{i}$, $\mathbf{M}_{i}^{'}$, and $\mathbf{M}_{i}^{''}$  have different initial values.
Table~\ref{tb_fix_acc} provides a proof-of-concept.
For each user, we randomly select one item from its local dataset and reverse the item's rating  to construct the dataset $\mathcal{D}_{i}^{j}$.
Once  $\mathbf{M}_{i}$, $\mathbf{M}_{i}^{'}$, $\mathbf{M}_{i}^{''}$ are trained,  we can infer the rating $r_{ij}$ in $\mathcal{D}_{i}$ 
 only based on the item's rating in $\mathcal{D}_{i}^{j}$ and the  distance of the item's embeddings in these three models.
 As shown in Table~\ref{tb_fix_acc},  the inference accuracy is higher than 90\% in most cases, showing the effectiveness of this inference attack method.  Based on this observation, if all other item ratings in $\mathcal{D}_{i}$ are known, we can infer $v_{j}$'s rating $r_{ij}$ by training $\mathbf{M}_{i}^{'}$ and $\mathbf{M}_{i}^{''}$ and then comparing their $v_{j}$'s item embedding distance with the uploaded parameters $\mathbf{V}^{t}_{i}$.

\begin{table}[!htbp]
  \setlength\tabcolsep{2.1pt}
  \caption{Accuracy of inferring randomly select items' ratings  for all users based on comparing Euclidean distances  $dist(\mathbf{v}_{j}, \mathbf{v}_{j}^{'})$ and $dist(\mathbf{v}_{j}, \mathbf{v}_{j}^{''})$.}\label{tb_fix_acc}
  \begin{tabular}{lccc}
  \hline
  \textbf{Models} & \multicolumn{1}{l}{\textbf{MovieLens-100K}} & \multicolumn{1}{l}{\textbf{Steam-200K}} & \multicolumn{1}{l}{\textbf{Amazon}} \\ \hline
  Fed-NCF          & 93.9\%                                      & 97.6\%                                  & 99.9\%                                         \\
  Fed-LightGCN     & 79.7\%                                      & 90.5\%                                  & 91.15\%                                        \\ \hline
  \end{tabular}
  \end{table}

However, the IMIA attacker does not know any item rating in $\mathcal{D}_{i}$, so the above  method cannot be directly used as the attack approach for FedRecs.
To implement IMIA attacks, we relax the requirement and generalize the observation: if most samples are the same on two datasets $\mathcal{D}$ and $\mathcal{D}^{'}$, and we train two models $\mathbf{M}$ and $\mathbf{M}^{'}$ on them respectively, the embeddings of counterpart items will be close if their ratings are the same.
Based on this assumption, when the server is curious about user $u_{i}$'s interaction data at epoch $t$, the server first randomly assigns ratings (i.e., 0 and 1) for each item in $\mathcal{V}_{i}$ according to the negative sampling ratio $\eta$.
For example, if $\eta$ is $1:4$, the server will randomly choose $25\%$ items as positive items and the remaining items as negative ones, thus constructing a fake dataset $\mathcal{D}_{i}^{fake}$.
Since the negative samples are empirically several times more than the positive items, $\mathcal{D}_{i}^{fake}$ and $\mathcal{D}_{i}$ still have a portion of common ratings. Still taking $\eta = 1:4$  as an example, although in the worst case  all positive items are wrongly assigned with rating 0, $\mathcal{D}_{i}^{fake}$ and $\mathcal{D}_{i}$ still have 50\% the same item ratings.
Then, the server trains a shadow model $\mathbf{M}_{i}^{fake}$ based on $\mathcal{D}_{i}^{fake}$.
After that, the malicious server calculates the  distance between item embeddings from $\mathbf{M}_{i}^{fake}$ and the uploaded item embeddings $V_{i}^{t}$, and it chooses $\gamma*\left|\mathcal{V}_{i}\right|$ items with the smallest distance as ``correct guess''.
The ratings of ``correct guess'' items  will be fixed in the next iteration.
Repeat the above steps several times until the server finishes inferring the positive item set $\hat{\mathcal{V}}_{i}^{+}$ for user $u_{i}$.
Since the whole inference attack process happens on the malicious server side by using uploaded public parameters, the client is unaware of the IMIA attack.
In addition, the malicious server can also store the target user's uploaded parameters and asynchronously execute the inference attack process without interrupting the normal training of FedRecs. Lines 23-32 in Alg.~\ref{alg_fedrec} describe the process of the proposed IMIA attack with pseudo-code.

\begin{algorithm}[!ht]
  \renewcommand{\algorithmicrequire}{\textbf{Input:}}
  \renewcommand{\algorithmicensure}{\textbf{Output:}}
  \caption{FedRec with IMIA attacker and defender.} \label{alg_fedrec}
  \begin{algorithmic}[1]
    \Require global epoch $T$; local epoch $L$; learning rate $lr$, negative sampling rate $\eta$, \dots;
    \Ensure global parameter $\mathbf{V}$, local client embedding $\mathbf{u}_{i}|_{i \in \mathcal{U}}$;
    \State Initializing global parameter $\mathbf{V}_{0}$;
    \For {each round t = 0, 1, ..., $T$}
        \State sampling a fraction of clients $\mathcal{U}_{t}$;
        \For{$u_{i}\in \mathcal{U}_{t}$}
          \State $\mathbf{V}_{i}^{t} \leftarrow$ \Call{ClientTrain}{$u_{i},\mathbf{V}_{t}, L$};
          \If {curious about $u_{i}$'s data}
            \State $\hat{\mathcal{V}}_{i}^{+}\leftarrow$\Call{Attacker}{$\mathbf{V}_{i}^{t}$,$\eta$, $\gamma$};
          \EndIf
        \EndFor
        \State $\mathbf{V}_{t+1} = \sum\limits_{u_{i} \in \mathcal{U}_{t}} \mathbf{V}_{i}^{t}$;
    \EndFor
    \Function{ClientTrain} {$u_{i},\mathbf{V}_{t}, L$}
      \State downloading $\mathbf{V}_{t}$ from the server;
      \State sampling negative items $\mathcal{V}_{i}^{neg}$;
      \If {use IMIA defender}
        \State $\mathbf{u}_{i}^{t+1}, \mathbf{V}_{i}^{t} \leftarrow$ training $L$ epochs with E.q.~\ref{eq_loss_with_defend};
      \Else
        \State $\mathbf{u}_{i}^{t+1}, \mathbf{V}_{i}^{t} \leftarrow$ training $L$ epochs with E.q.~\ref{eq_ori_loss};
      \EndIf
      \State if use LDP, add noise with E.q.~\ref{eq_ldp};
      \State \Return $\mathbf{V}_{i}^{t}$
    \EndFunction
    \Function{Attacker}{$\mathbf{V}_{i}^{t}$, $\eta$, $\gamma$}
      \State $\hat{\mathcal{V}}_{i}^{+}=\{\}$
      \State $\mathcal{V}_{i}\leftarrow$ select updated items according to $\mathbf{V}_{i}^{t}$ and $\mathbf{V}_{t}$;
      \While {$\left|\hat{\mathcal{V}}_{i}^{+}\right| <\eta\left|\mathcal{V}_{i}\right|$}
        \State randomly assign ratings to $v_{j} \in \mathcal{V}_{i} \setminus \hat{\mathcal{V}}_{i}^{+}$;
        \State train fake model $\mathbf{M}_{i}^{fake}$ on constructed dataset;
        \State $\hat{\mathcal{V}}_{i}^{+}, \hat{\mathcal{V}}_{i}^{-}\leftarrow$ select $\gamma*\left|\mathcal{V}_{i}\right|$ items using $dist(\mathbf{V}_{i}^{t}, \mathbf{V}_{i}^{fake})$;
      \EndWhile
      \State \Return $\hat{\mathcal{V}}_{i}^{+}$
    \EndFunction
    \end{algorithmic}
\end{algorithm}

\subsection{Interaction-level Membership Inference Defender}~\label{sec_imia_def}
In Section~\ref{sec_vanilla} and \ref{sec_ldp_against}, the experimental results demonstrate that both vanilla FedRecs and FedRecs with LDP are  vulnerable to the new attack IMIA, highlighting the need for
 a new defense mechanism.  The experimental results in Table~\ref{tb_vanilla_atk} and~\ref{tb_ldp_imia} show that  Fed-LightGCN is more resistant to IMIA.  This may be because the private user embeddings in Fed-LightGCN learn more useful information and patterns than in Fed-NCF. 
  Since the private user embeddings in Fed-LightGCN capture more user-item interaction patterns,  it is harder for the curious server to infer interactions only from public parameters.

\begin{figure}[!htbp]
  \centering
  \subfloat[Deviation trend in Fed-NCF.]{\includegraphics[width=1.67in]{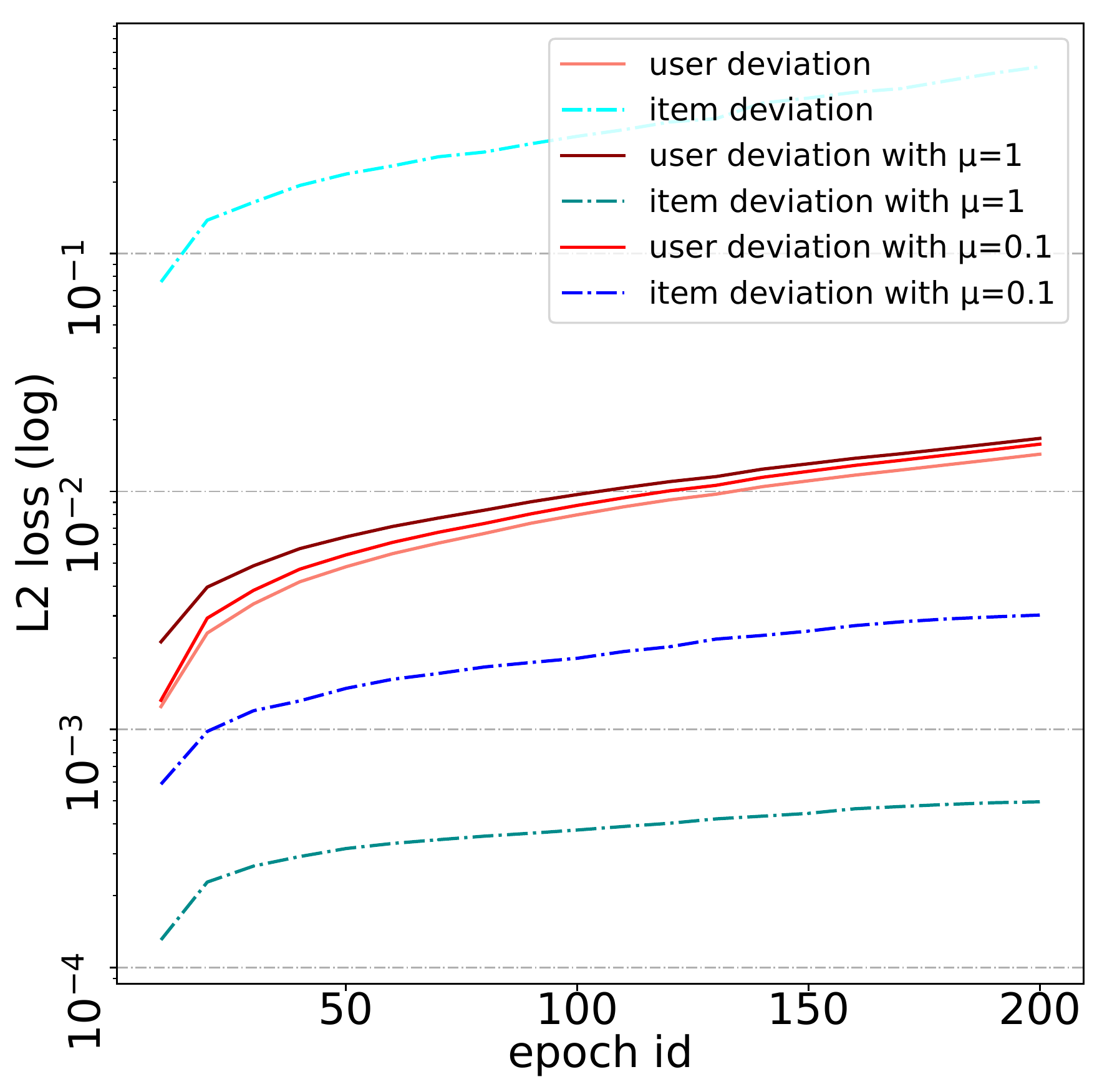}\label{fig_ncf_deviation}}
  \subfloat[Deviation trend in Fed-LightGCN.]{\includegraphics[width=1.67in]{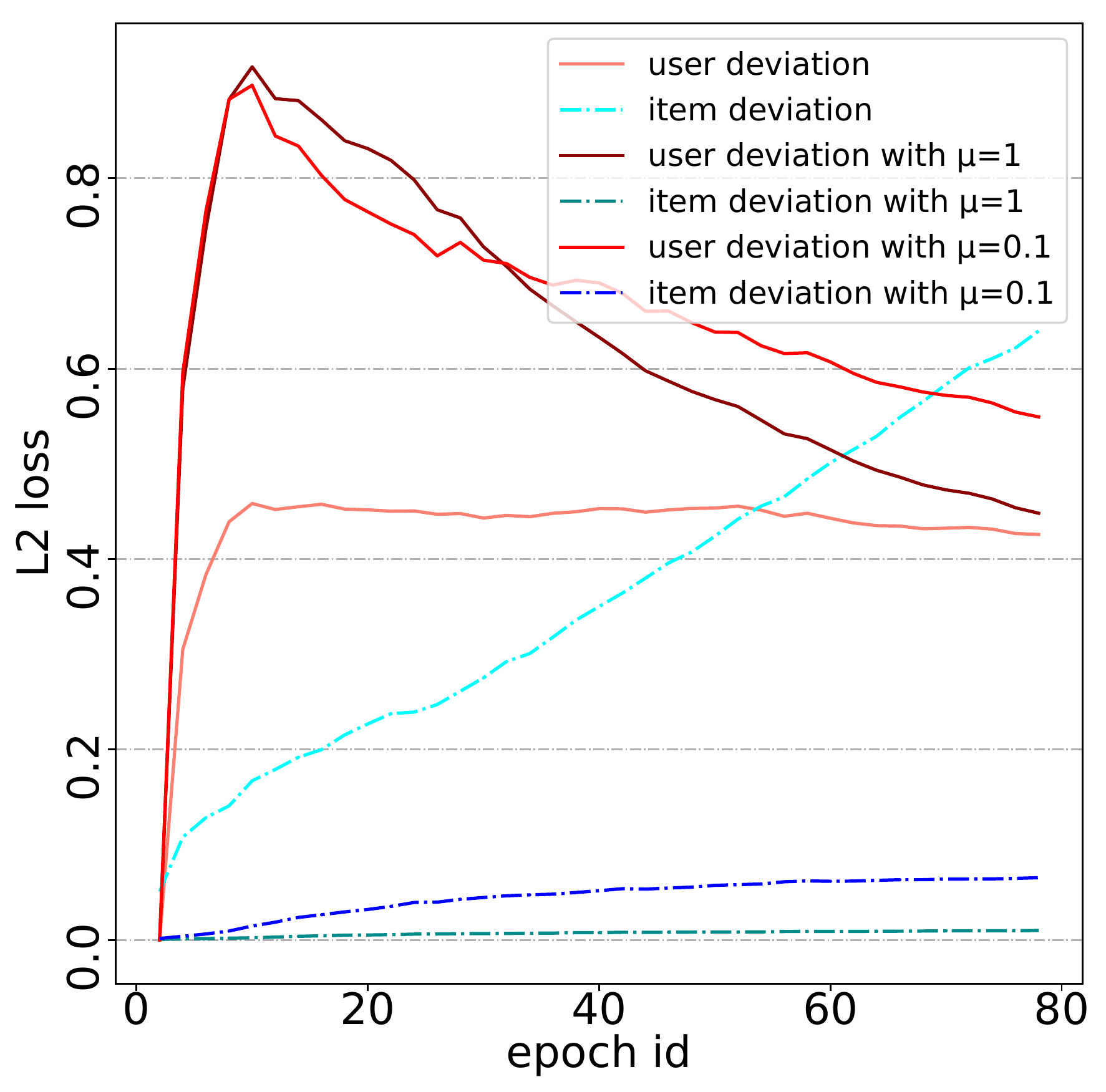}\label{fig_lightgcn_deviation}}
  \caption{Trend of embedding deviation over time until convergence in Fed-NCF and Fed-LightGCN on MovieLens-100K.}\label{fig_deviation}
\end{figure}

To further validate our hypothesis, we compare the deviation of user/item embeddings in the training process from their initial values using L2 loss (i.e., $dist^{2}(v^{t}_{i}-v^{0}_{i})$).   Fig.~\ref{fig_deviation} illustrates the trend of the average deviation over training time.
In Fig.~\ref{fig_deviation}, the deviation of item embeddings is much larger than user embeddings' deviation.
In other words, on average, user embeddings do not  change as much as item embeddings during the whole training process, therefore user embeddings learn less information and patterns. 
Further, by comparing Fig.~\ref{fig_ncf_deviation} and Fig.~\ref{fig_lightgcn_deviation}, we can see that user embeddings in Fed-NCF vary
 much less than in Fed-LightGCN, which supports our hypothesis.
Note that for the sake of visualization, we log the L2 loss value in Fig.~\ref{fig_ncf_deviation} because of the large difference between user and item embedding deviation.

Motivated by the above observation, we propose a novel IMIA defender.
 The basic idea of LDP is to add noise to the shared parameters to distort the sensitive information behind the shared parameters, leading to catastrophic performance dropping.
Unlike LPD, the key idea of our defender is to restrict the learning ability of public parameters so that they will convey less information to the curious central server. To implement that, we add a constraint term in the original FedRec loss function E.q.~\ref{eq_ori_loss}, as follows:
\begin{equation}\label{eq_loss_with_defend}
  \mathcal{L} = \mathcal{L}^{rec} + \mu\left\| \mathbf{V}_{i}^{t}-\mathbf{V}_{t} \right\|
\end{equation}
The constraint term limits the update of the public parameters $\mathbf{V}_{t}$ on each local client/device.  Consequently, to optimize $\mathcal{L}^{rec}$, the recommender model would enforce the private embeddings to learn more information and patterns.
Fig.~\ref{fig_deviation} shows the embedding deviation trend after applying our defender to Fed-NCF and Fed-LightGCN.
User embedding deviation becomes larger than vanilla FedRecs, while item embedding deviation significantly drops.
More details of embedding deviation are in Appendix~\ref{app_deviation}.

\section{Experiments}
\subsection{Datasets}
We use three real-world datasets (MovieLens-100K~\cite{harper2015movielens}, Steam-200K~\cite{cheuque2019recommender}, and Amazon Cell Phone~\cite{he2016ups})  from various domains (movie recommendation, game recommendation, and cell phone recommendation) to evaluate the performance of our IMIA attacker and defender.
The statistics of these datasets are shown in Table~\ref{tb_dataset}.
MovieLens-100K contains $100,000$ interactions between $943$ users and $1,682$ items.
There are  $3,753$ users, $5,134$ items, and $114,713$ interactions in Steam-200K.
Amazon Cell Phone consists of  $13,174$ users,  $5,970$ cell phone related items, and $103,593$ interactions.
Note that the densities of these three datasets are different.
MovieLens-100K is the densest dataset, while Amazon Cell Phone is the most sparse one.
Following~\cite{zhang2022pipattack}, we binarize the user feedback, where all ratings are transformed to $r_{ij}=1$ and negative instances are sampled with $1:4$ ratio.
Besides, we utilize the leave-one-out method to split the training, validation, and test sets.  

\begin{table}[!ht]
  \centering
  \setlength\tabcolsep{2.pt}
  \caption{Statistics of recommendation datasets}\label{tb_dataset}
  \begin{tabular}{lccccc}
  \hline
  \textbf{Dataset}  & \multicolumn{1}{l}{\textbf{\#users}} & \multicolumn{1}{l}{\textbf{\#items}} & \multicolumn{1}{l}{\textbf{\#interactions}} & \multicolumn{1}{l}{\textbf{Avg.}} & \multicolumn{1}{l}{\textbf{Density}} \\ \hline
  MovieLens-100K    & 943                                  & 1,682                                 & 100,000                                     & 106                                      & 6.30\%                               \\
  Steam-200K        & 3,753                                 & 5,134                                 & 114,713                                     & 31                                       & 0.59\%                               \\
  Amazon  & 13,174                               & 5,970                                & 103,593                                     & 8                                        & 0.13\%                               \\ \hline
  \end{tabular}
  \end{table}

\subsection{Evaluation Metrics}
To measure the effectiveness of IMIA attackers, we employ the widely used classification metric F1 score to evaluate inference performance.
To evaluate the recommendation performance, we adopt the widely used hit ratio at rank $10$ (Hit@10), which measures the ratio of ground truth items that appear in the top-10 recommendation list.
\subsection{Baselines}
Since none of the prior works conducts interaction-level membership attacks on FedRecs,  we design two baselines.

\textbf{Random Attack.} For each client $u_{i}$, the server randomly selects a group of items from $\mathcal{V}_{i}$ as the positive items based on the negative sampling ratio $\eta$.
Comparing with Random Attack can reveal whether a privacy issue of user interaction data exists.

\textbf{K-means Attack.} Since we do not have any labels of user-item interaction samples, IMIA can naturally be treated as a clustering problem.
We adopt K-means~\cite{hartigan1979algorithm} algorithm to divide items into two clusters based on the client's uploaded public parameters $\mathbf{V}_{i}^{t}$. Positive items are chosen from the cluster with lower SSE (the sum of squared errors).
The intuition of K-means Attack is that for a user, the positive items are more similar to each other than diverse negative items due to the coherence principle of personal interests, therefore, their embeddings will also be more coherent.

\subsection{Parameter Settings}
For both Fed-NCF and Fed-LightGCN, the dimension of user and item embeddings is $64$, and $3$ neural layers with dimensions $128, 64, 32$ are used to process the concatenated user and item embedding.
The negative sampling ratio $\eta$ is set to $1:4$, as this ratio can well balance the training effectiveness and efficiency for most pair-wise loss functions and has been widely used.
The local training batch size and local epoch size are $64$ and $20$, respectively.
Adam~\cite{kingma2014adam} optimizer with $0.001$ learning rate is employed to optimize local models.
To ensure the model convergence, the maximum global epoch is set to $200$. $\gamma$ is set to $20\%$.
 We also perform the sensitivity  analysis of key hyper-parameters in the experiment.

\subsection{Performance of IMIA Attackers}\label{sec_vanilla}
\begin{figure}[!ht]
    \centering
    \includegraphics[width=0.4\textwidth]{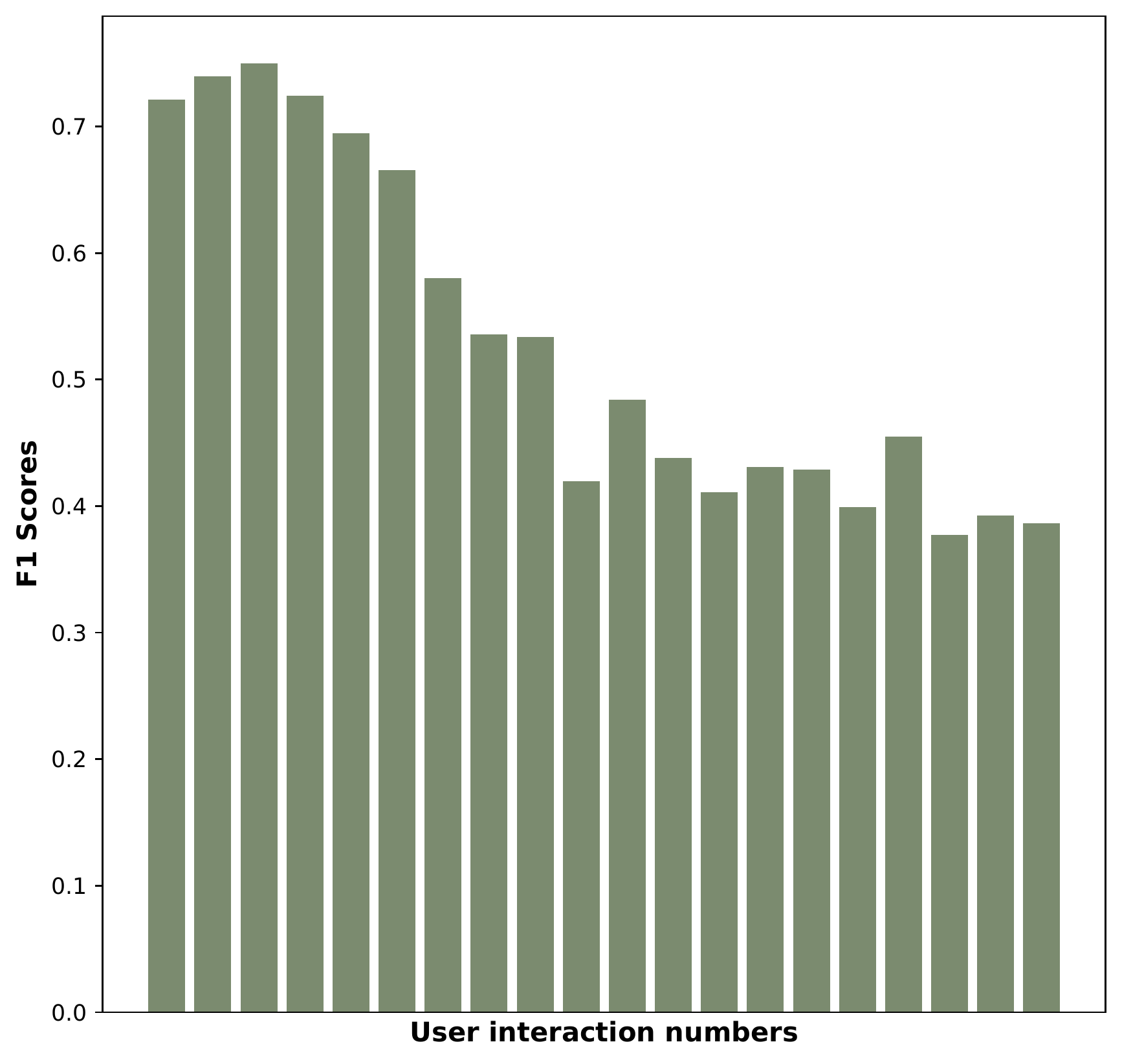}
    \caption{IMIA attacker performance for users with different number of interactions on MovieLens-100K.}\label{fig_user_interaction}
  \end{figure}

Table~\ref{tb_vanilla_atk} presents three attackers' performances on two FedRecs and three datasets.
The results are average F1 scores that reflect the inference effectiveness of the IMIA attacker. The results in Table~\ref{tb_vanilla_atk} highlight that \textbf{vanilla FedRecs have a high risk of user interaction data leakage}, since the performance of our IMIA attacker is much better than Random Attack.
Besides, comparing K-means and our attacker, we can see that the naive clustering method cannot effectively infer user interaction information.
Furthermore, by comparing our IMIA attacker's performances crossing datasets, we can find that FedRecs trained on Steam-200K and Amazon Cell Phone are more vulnerable to IMIA than the ones trained on MovieLens-100K.
With the statistics of datasets in Table~\ref{tb_dataset}, we believe that this phenomenon is related to the number of user interactions  because the average number of user interactions on MovieLens-100K is much higher than that on the other two datasets.
To further investigate this phenomenon on MovieLens-100K, we cluster users into 20 groups  according to their interaction numbers and report their average F1 score in Fig.~\ref{fig_user_interaction}.    
The results show that \textbf{users with fewer interactions have a higher risk of interaction data leakage}.
Appendix~\ref{app_interaction} analyzes this phenomenon on all datasets.

\begin{table}[!htbp]
  \centering
  \setlength\tabcolsep{2.1pt}
  \caption{The performance (F1 scores) of attackers on vanilla FedRecs. ML-100K is short for MovieLens-100K, Amazon is short for Amazon Cell Phone.}~\label{tb_vanilla_atk}
  \begin{tabular}{l|l|lll}
  \hline
  \textbf{Model}                         & \textbf{Attack} & \textbf{ML-100K} & \textbf{Steam-200K} & \textbf{Amazon} \\ \hline
                                         & \textbf{Random} & {\ul 0.2079}            & {\ul 0.2019}        & {\ul 0.1998}    \\ \hline
  \multirow{2}{*}{\textbf{Fed-NCF}}      & \textbf{K-means} & 0.3183                  & 0.2477              & 0.2458          \\
                                         & \textbf{Ours}   & \textbf{0.5928}         & \textbf{0.6707}     & \textbf{0.6516} \\ \hline
  \multirow{2}{*}{\textbf{Fed-LightGCN}} & \textbf{K-means} & 0.1460                  & 0.2573              & 0.2697          \\
                                         & \textbf{Ours}   & \textbf{0.3900}         & \textbf{0.6007}     & \textbf{0.4328} \\ \hline
  \end{tabular}
  \end{table}

  \begin{table*}[!htbp]
    \centering
    \caption{The result of Local Differential Privacy (LDP) against our IMIA attacker. F1 is the attacker's performance, and the lower scores ($\downarrow$) are better. Hit@10 ($\uparrow$) measures recommendation performance, and the higher scores are better.}\label{tb_ldp_imia}
    \begin{tabular}{ll|cccccccc}
    \hline
    \textbf{Model}                         & \textbf{Dataset}    & \multicolumn{8}{c}{\textbf{Noise Scale}}                                                                                                                                                                                                                                                      \\ \hline
                                           &                     & \multicolumn{2}{c}{$\lambda=$\textbf{0.0}}                                      & \multicolumn{2}{c}{$\lambda=$\textbf{0.001}}                                    & \multicolumn{2}{c}{$\lambda=$\textbf{0.01}}                                     & \multicolumn{2}{c}{$\lambda=$\textbf{0.1}}                                      \\ \cline{3-10} 
                                           &                     & \multicolumn{1}{c}{\textbf{F1}$\downarrow$} & \multicolumn{1}{c}{\textbf{Hit@10}$\uparrow$} & \multicolumn{1}{c}{\textbf{F1}$\downarrow$} & \multicolumn{1}{c}{\textbf{Hit@10}$\uparrow$} & \multicolumn{1}{c}{\textbf{F1$\downarrow$}} & \multicolumn{1}{c}{\textbf{Hit@10}$\uparrow$} & \multicolumn{1}{c}{\textbf{F1}$\downarrow$} & \multicolumn{1}{c}{\textbf{Hit@10}$\uparrow$} \\ \hline
    \multirow{3}{*}{\textbf{Fed-NCF}}       & \textbf{ML-100K}    & {\ul 0.5928}                          & \textbf{0.3690}                     & 0.5474                          & 0.3308                              & 0.3954                          & 0.2958                              & \textbf{0.2520}                 & 0.1696                              \\
                                           & \textbf{Steam-200K} & {\ul 0.6707}                          & \textbf{0.6645}                     & 0.6012                          & 0.5901                              & 0.3334                          & 0.4524                              & \textbf{0.2199}                 & 0.2224                              \\
                                           & \textbf{Amazon}     & {\ul 0.6516}                          & \textbf{0.2176}                     & 0.6260                          & 0.1984                              & 0.2933                          & 0.1505                              & \textbf{0.2126}                 & 0.1217                              \\ \hline
    \multirow{3}{*}{\textbf{Fed-LightGCN}} & \textbf{ML-100K}    & {\ul 0.3900}                          & \textbf{0.4072}                     & 0.3786                          & 0.3923                              & 0.2816                          & 0.3658                              & \textbf{0.2357}                 & 0.3138                              \\
                                           & \textbf{Steam-200K} & {\ul 0.6007}                          & \textbf{0.6943}                     & 0.5690                          & 0.6957                              & 0.3392                          & 0.6890                              & \textbf{0.2188}                 & 0.5123                              \\
                                           & \textbf{Amazon}     & {\ul 0.4328}                          & \textbf{0.1796}                     & 0.3483                          & 0.1717                              & 0.2642                          & 0.1720                              & \textbf{0.2209}                 & 0.1562                              \\ \hline
    \end{tabular}
    \end{table*}
    \begin{table*}[!htbp]
      \caption{The result of our defender against IMIA. The best results on each dataset are bold.}\label{tb_our_defend}
      \resizebox{\textwidth}{!}{
      \begin{tabular}{ll|cccccccccc}
      \hline
      \textbf{Model}                         & \textbf{Dataset}    & \multicolumn{10}{c}{\textbf{Constraint Scale}}                                                                                                                                                                                                                                                                                                                        \\ \hline
                                             &                     & \multicolumn{2}{c}{$\mu=$\textbf{0.0}}                                      & \multicolumn{2}{c}{$\mu=$\textbf{0.1}}                                      & \multicolumn{2}{c}{$\mu=$\textbf{0.4}}                                      & \multicolumn{2}{c}{$\mu=$\textbf{0.7}}                                      & \multicolumn{2}{c}{$\mu=$\textbf{1.0}}                                               \\ \cline{3-12} 
                                             &                     & \multicolumn{1}{c}{\textbf{F1}$\downarrow$} & \multicolumn{1}{c}{\textbf{Hit@10}$\uparrow$} & \multicolumn{1}{c}{\textbf{F1}$\downarrow$} & \multicolumn{1}{l}{\textbf{Hit@10}$\uparrow$} & \multicolumn{1}{l}{\textbf{F1}$\downarrow$} & \multicolumn{1}{l}{\textbf{Hit@10}$\uparrow$} & \multicolumn{1}{l}{\textbf{F1}$\downarrow$} & \multicolumn{1}{l}{\textbf{Hit@10}$\uparrow$} & \multicolumn{1}{l}{\textbf{F1}$\downarrow$} & \multicolumn{1}{l}{\textbf{Hit@10}$\uparrow$} \\ \hline
      \multirow{3}{*}{\textbf{Fed-NCF}}       & \textbf{ML-100K}    & {\ul 0.5928}                    & 0.3690                              & 0.2638                          & 0.3605                              & \textbf{0.2140}                 & \textbf{0.3743}                     & 0.2166                          & 0.3563                              & 0.2145                          & 0.3531                              \\
                                             & \textbf{Steam-200K} & {\ul 0.6707}                    & \textbf{0.6645}                     & 0.3888                          & 0.6005                              & 0.2667                          & 0.6011                              & 0.2213                          & 0.5960                              & \textbf{0.2058}                 & 0.5960                              \\
                                             & \textbf{Amazon}     & {\ul 0.6516}                    & \textbf{0.2176}                     & 0.4761                          & 0.2142                              & 0.3368                          & 0.2129                              & \textbf{0.3079}                 & 0.2126                              & 0.3240                          & 0.2121                              \\ \hline
      \multirow{3}{*}{\textbf{Fed-LightGCN}} & \textbf{ML-100K}    & {\ul 0.3900}                    & 0.4072                              & 0.2130                          & \textbf{0.4082}                     & 0.1892                          & 0.3891                              & 0.1811                          & 0.3796                              & \textbf{0.1741}                 & 0.3870                              \\
                                             & \textbf{Steam-200K} & {\ul 0.6007}                    & \textbf{0.6943}                     & 0.4730                          & 0.6584                              & 0.4620                          & 0.5830                              & 0.4205                          & 0.5582                              & \textbf{0.2246}                 & 0.5472                              \\
                                             & \textbf{Amazon}     & {\ul 0.4328}                    & 0.1796                              & \textbf{0.2281}                 & \textbf{0.1920}                     & 0.2847                          & 0.1821                              & 0.3231                          & 0.1704                              & 0.3308                          & 0.1615                              \\ \hline
      \end{tabular}}
      \end{table*}
    
    \begin{table}[!htbp]
      \setlength\tabcolsep{1.8pt}
      \centering
      \caption{Comparison of $\frac{\left|\Delta F1\right|}{\left|\Delta Hit@10\right|}$ for LDP and our defender. Higher scores represent the more cost-effective defense. NCF and LightGCN are short for ``Fed-NCF'' and ``Fed-LightGCN''.}\label{tb_delta}
      \begin{tabular}{l|cccccc}
      \hline
      \textbf{Defense} & \multicolumn{2}{c}{\textbf{ML-100K}}                                             & \multicolumn{2}{c}{\textbf{Steam-200K}}                                          & \multicolumn{2}{c}{\textbf{Amazon}}                                              \\ \hline
                        & \multicolumn{1}{l}{\textbf{NCF}} & \multicolumn{1}{l}{\textbf{LightGCN}} & \multicolumn{1}{l}{\textbf{NCF}} & \multicolumn{1}{l}{\textbf{LightGCN}} & \multicolumn{1}{l}{\textbf{NCF}} & \multicolumn{1}{l}{\textbf{LightGCN}} \\ \hline
      \textbf{LDP}     & 1.70                                 & 1.65                                      & 1.01                                 & 2.09                             & 4.57                                 & 9.05                                      \\
      \textbf{ours}    & \textbf{71.47}                       & \textbf{10.68}                            & \textbf{6.78}                        & \textbf{2.55}                                      & \textbf{68.74}                       & \textbf{16.50}                            \\ \hline
      \end{tabular}
      \end{table}
  
Finally, the comparison of IMIA attackers' performances on Fed-NCF and Fed-LightGCN shows that \textbf{Fed-LightGCN is more resistant to IMIA than Fed-NCF}.
This may be because that \textbf{private parameters (i.e., user embeddings) in Fed-LightGCN  learn more useful information than in Fed-NCF}, since user embeddings in Fed-LightGCN aggregate information from  item embedding via convolution operation.
As a result, only using public parameters to infer user interaction records  becomes harder.
In Appendix~\ref{app_deviation}, we further show the embeddings' deviation from their initial values. 
The results support our explanation.
The above observation motivates us to design our effective IMIA defender (see Section~\ref{sec_imia_def}), which attempts to limit the learning ability of public parameters and enforce private parameters to learn more patterns.

\subsection{Effectiveness of LDP Against IMIA}\label{sec_ldp_against}
As the most classical and widely used privacy-preserving approach,
LDP can effectively prevent attribute inference attacks on FedRecs~\cite{zhang2022comprehensive}.
Here, we conduct this experiment to study whether LDP can defend against the new inference attack IMIA. 
Table~\ref{tb_ldp_imia} presents the results of LDP with different noise scales against IMIA attacks.
$\lambda=0.0$ means FedRecs without LDP.
The results indicate that with subtle noise (e.g. $\lambda=0.001$), LDP cannot well protect user interaction data.
Adding more noises (e.g. $\lambda=0.1$) can defend against our IMIA attacker, however, stronger noises severely degenerate the recommendation performance of FedRecs.

To measure how much recommendation performance LDP needs to sacrifice to effectively defend the attacker, we calculate $\frac{\left|\Delta F1\right|}{\left|\Delta Hit@10\right|}$ for the LDP which degenerates the IMIA attacker's performance to the level of Random Attack.
Intuitively, $\frac{\left|\Delta F1\right|}{\left|\Delta Hit@10\right|}$ measures the change ratio of the attacker's performance and recommendation performance.
Lower scores represent that the defender has to sacrifice more recommendation performance to reduce the attacker's threat.
Table~\ref{tb_delta} shows that LDP would sacrifice too much recommendation performance to alleviate IMIA threats.
As a result, \textbf{LDP is not cost-effective to defend against IMIA}.

\subsection{Effectiveness of IMIA Defender}
Since LDP cannot effectively mitigate IMIA threats, we propose a novel defense mechanism against the IMIA attack.
The results of our defender against IMIA are shown in Table~\ref{tb_our_defend} where we vary the values of the hyper-parameter $\mu$ from 0.0 to 1.0, and $\mu=0.0$ represents the vanilla FedRecs.
With our defense method, the attacker's performance is reduced to the level of random guesses in all cases.
Meanwhile, the recommender's performance is even improved in some cases (e.g., Fed-NCF on ML-100K, Fed-LightGCN on ML-100K, and Amazon Cell Phone) due to the regularization effect of the constraint term in the loss function, which indicates that \textbf{when restricting the updates of public parameters, the recommendation models can still  achieve good recommendation performance by enforcing private parameters to learn more patterns}.

Table~\ref{tb_delta} shows the comparison  between LDP and our defender.
The higher scores represent that the defender invalids the IMIA attacker with less performance loss.
As we can see, in all cases, our defender is more cost-effective than LDP.
Specifically, our defender's $\frac{\left|\Delta F1\right|}{\left|\Delta Hit@10\right|}$ scores for Fed-NCF on MovieLens-100K and Amazon Cell Phone are nearly $40$ times and $15$ times higher than LDP.
In conclusion, our defender provides a more cost-effective solution against IMIA than LDP.

\subsection{Attack with More Prior Knowledge}
\begin{figure}[!ht]
  \centering
  \includegraphics[width=0.35\textwidth]{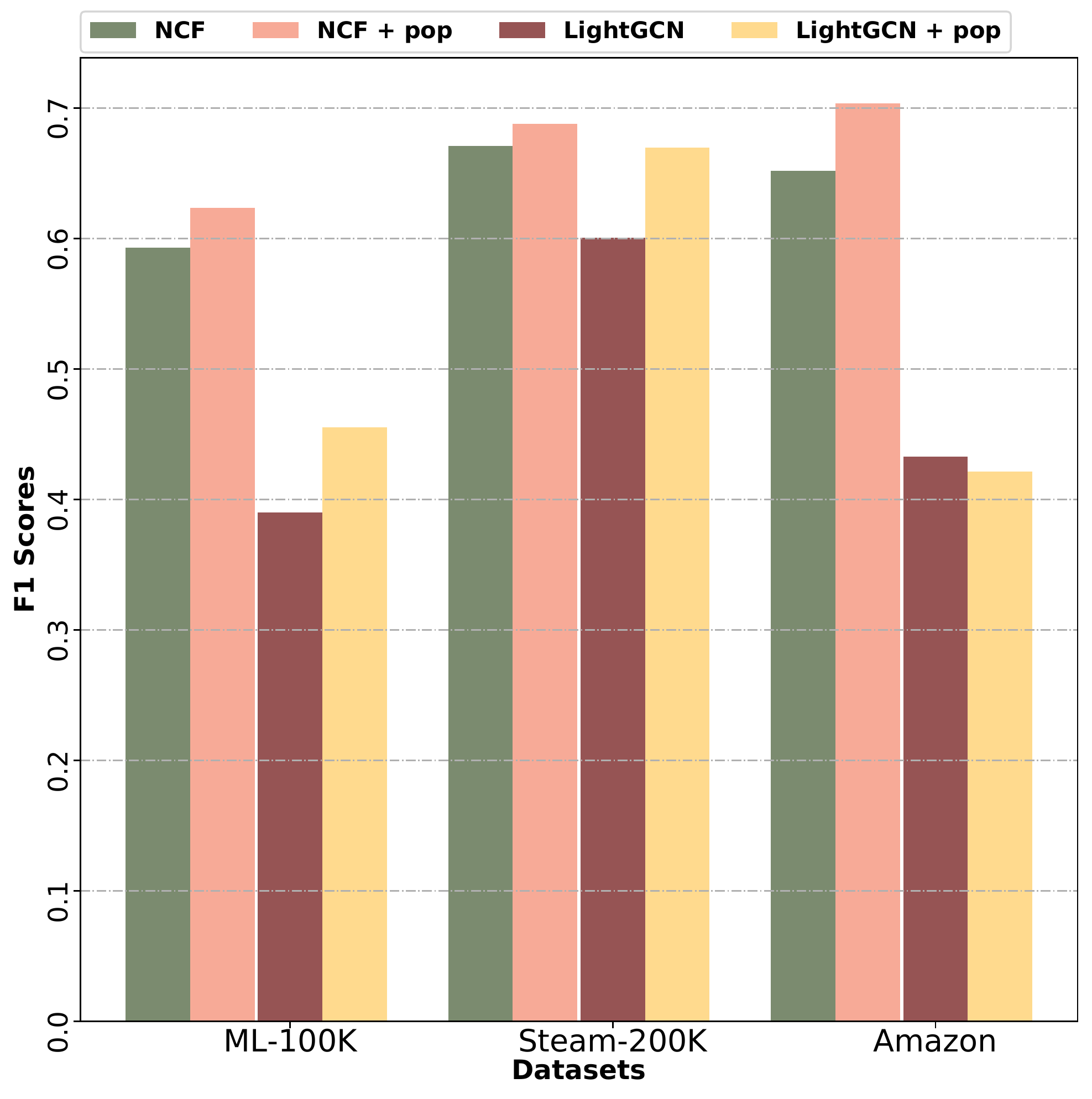}
  \caption{IMIA with popularity information. NCF and LightGCN are short for ``IMIA for Fed-NCF'' and ``IMIA for Fed-LightGCN''. ``pop'' means popularity information.}\label{fig_popularity}
\end{figure}

As mentioned in Section~\ref{sec_imia_attack}, to make the threat more realistic, we strictly restrict the curious server's prior knowledge with only uploaded parameters and some hyper-parameters such as learning rate and sampling ratio.
In this section, we explore one possible prior knowledge that the server may have chances to access: the popularity information of items. Although the popularity information is not always accessible, it is still available in many scenarios. In this part, we assume that the server knows the top 10\% popular items. Based on the popularity information, instead of randomly assigning ratings to items at the initial phase, the server assigns positive ratings to popular items with a higher probability. 
 Fig.~\ref{fig_popularity} shows that with the item popularity information, the IMIA attacker's performance is improved in most cases.

\section{Related Work}
In this section, we mainly introduce the related works of attacks against federated learning and attacks against federated recommender systems. 
The recent progress of recommender systems, federated recommender systems, federated learning, and local differential privacy can be referred to~\citep{zhang2019deep,yang2020federated,lyu2020threats,trung2020adaptive,wang2020next}.

\subsection{Attack against Federated Learning}
Recently, varieties of attacks were proposed to access privacy risks in federated learning (FL)~\cite{lyu2020threats,rodriguez2022survey}.
These attacks include threats such as model inversion~\cite{zhang2020secret}, attribute inference~\cite{ganju2018property}, and membership inference.
In this paper, we mainly discuss membership inference attacks.
Nasr et al.~\cite{nasr2019comprehensive} took the first comprehensive study of class-level membership inference attack in FL under both white-box and black-box settings.
Then, many works took further steps to study more fine-grained membership inference attacks, e.g.~\cite{nguyen2017argument,wang2019beyond,zhao2021user,hu2021source,suri2022subject}.
However, existing membership inference attacks cannot be used in FedRec because of the major differences mentioned in Section~\ref{sec:introduction}.

\subsection{Attack against Federated Recommendation}
Zhang et al.~\cite{zhang2022comprehensive} conducted the first analysis of FedRec's privacy-preserving, however, their work only reveals attribute-level leakage risks.
Some research discussed the user rating privacy issue of FedRec with explicit feedback~\cite{lin2020fedrec,liang2021fedrec++,lin2021fr,lin2022generic}, but the interaction privacy issue of FedRec with implicit feedback is another pair of shoes.
Other attack methods~\cite{zhang2022pipattack} aim to promote/demote item's rank, which cannot reveal the privacy issue of FedRecs.
As a result, the privacy issue of FedRecs is still under explored.
Besides, the defense method for improving federated recommendation's privacy protection is also under explored~\cite{yuan2022federated}.

\section{Conclusion}
In this paper, we perform the first study of interaction-level membership inference attacks (IMIA) in federated recommender systems (FedRecs) to reveal the privacy issue of user-item interactions.
We first design an attacker from the curious-but-honest server side.
The attacker infers the target user's private interaction based on its uploaded public parameters by iteratively training shadow models on shadow datasets.
We implement IMIA attack with two commonly used FedRecs on three real-world datasets.
The experimental results validate the threats of IMIA for FedRecs.
Furthermore, we find that the classical privacy-preserving method, LDP, cannot effectively defend against our attack.
In light of this, we propose a novel defender to mitigate IMIA threats with imperceptible influence on the recommendation performance.

\begin{acks}
  This work is supported by Australian Research Council Future Fellowship (Grant No. FT210100624), Discovery Project (Grant No. DP190101985), and Discovery Early Career Research Award (Grant No. DE200101465).
\end{acks}

\bibliographystyle{ACM-Reference-Format}
\balance
\bibliography{sample-base}

\appendix
\begin{table*}[ht]
  \centering
  \caption{The average deviation (L2 loss) of embeddings from initial point to the converged model point.}\label{tb_deviation}
  \begin{tabular}{l|c|cccccc}
  \hline
  \multirow{2}{*}{}                        & \multicolumn{1}{l|}{} & \multicolumn{2}{c}{\textbf{ML-100K}}     & \multicolumn{2}{c}{\textbf{Steam-200K}}  & \multicolumn{2}{c}{\textbf{Amazon}}      \\ \cline{2-8} 
                                           & $\mu$            & \textbf{Fed-NCF} & \textbf{Fed-LightGCN} & \textbf{Fed-NCF} & \textbf{Fed-LightGCN} & \textbf{Fed-NCF} & \textbf{Fed-LightGCN} \\ \hline
  \multirow{3}{*}{\textbf{User Embedding}} & 0.0                   & 0.0143           & 0.4258                & 0.0884           & 0.2666                & 0.0994           & 0.2252                \\
                                           & 0.1                   & 0.0816           & 0.5493                & 0.0932           & 0.3408                & 0.0996           & 0.2260                \\
                                           & 1.0                   & 0.1580           & 0.4481                & 0.0935           & 0.2935                & 0.0994           & 0.2031                \\ \hline
  \multirow{3}{*}{\textbf{Item Embedding}} & 0.0                   & 0.6088           & 0.6396                & 0.3144           & 0.2231                & 0.0667           & 0.0717                \\
                                           & 0.1                   & 0.0030           & 0.0653                & 0.0042           & 0.0303                & 0.0060           & 0.0313                \\
                                           & 1.0                   & 0.0004           & 0.0099                & 0.0005           & 0.0058                & 0.0009           & 0.0081                \\ \hline
  \end{tabular}
  \end{table*}
\begin{figure*}[ht]
  \centering
  \subfloat[F1 on MovieLens-100K with Fed-NCF.]{\includegraphics[width=1.67in]{user_interaction_ml_ncf.pdf}\label{fig_ml_ncf_interaction}}
  \hfil
  \subfloat[F1 on MovieLens-100K with Fed-LightGCN.]{\includegraphics[width=1.67in]{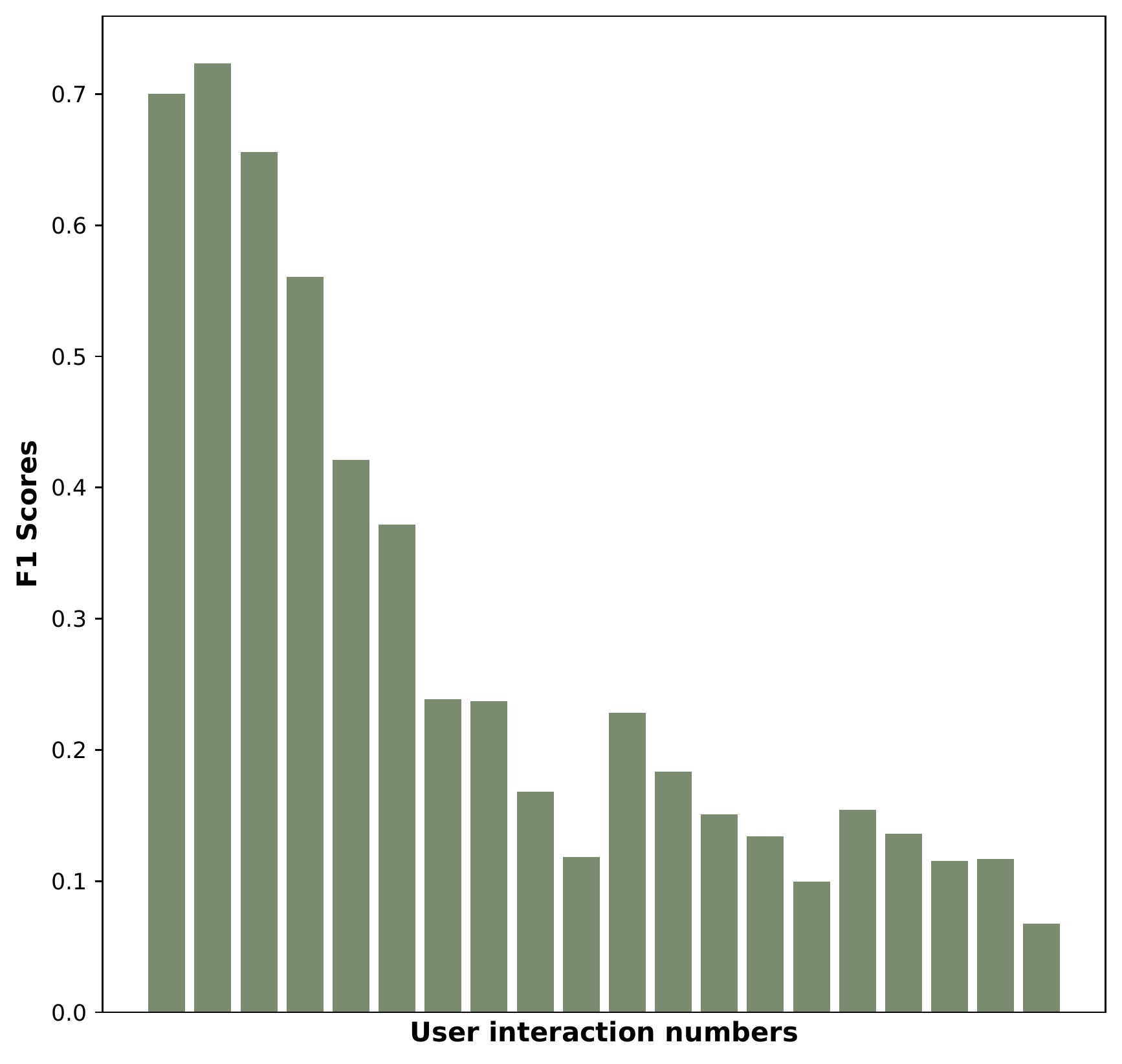}\label{fig_ml_lightgcn_interaction}}
  \hfil
  \subfloat[F1 on Steam-200K with Fed-NCF.]{\includegraphics[width=1.67in]{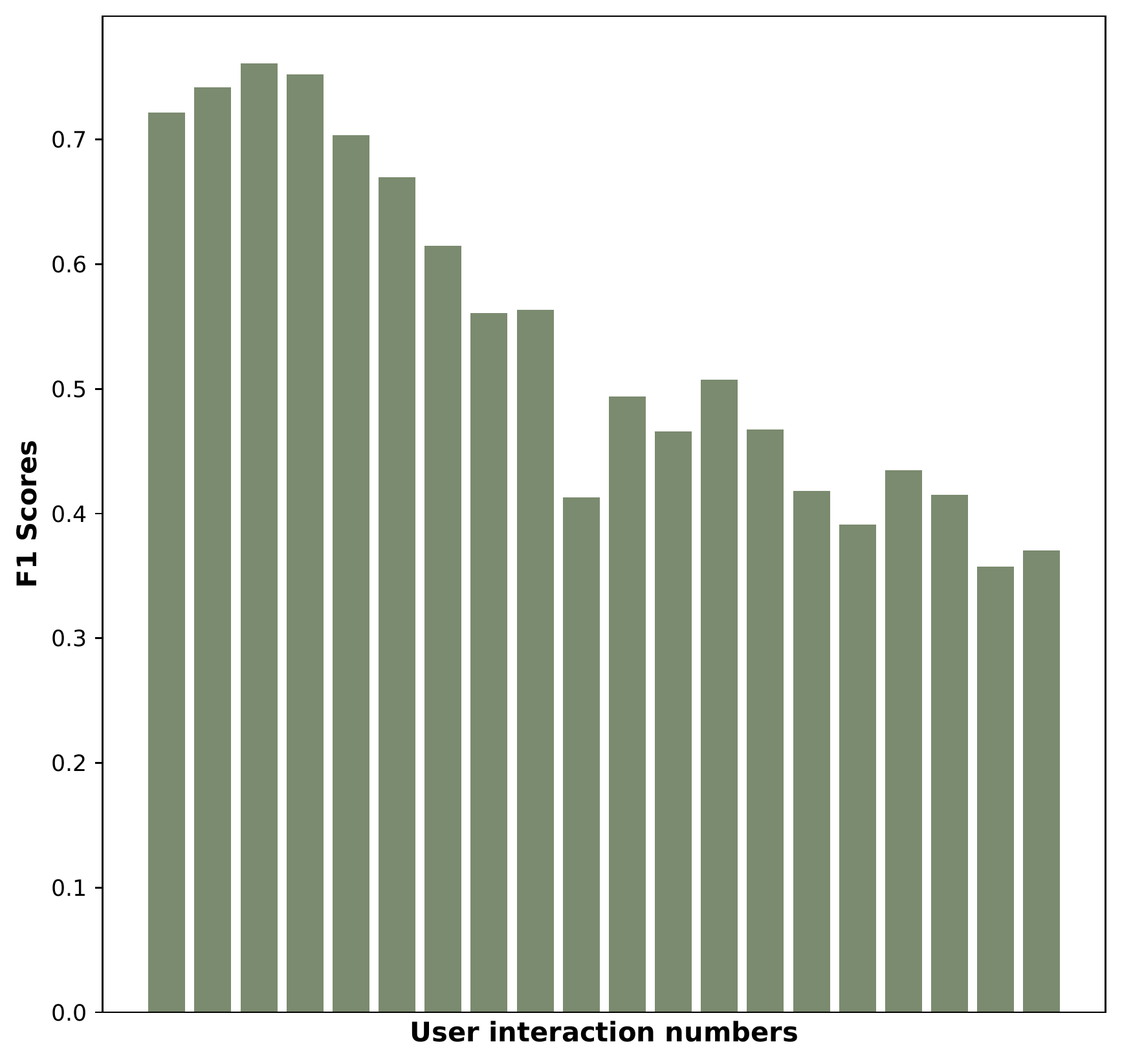}\label{fig_steam_ncf_interaction}}
  \hfil
  \subfloat[F1 on Steam-200K with Fed-LightGCN.]{\includegraphics[width=1.67in]{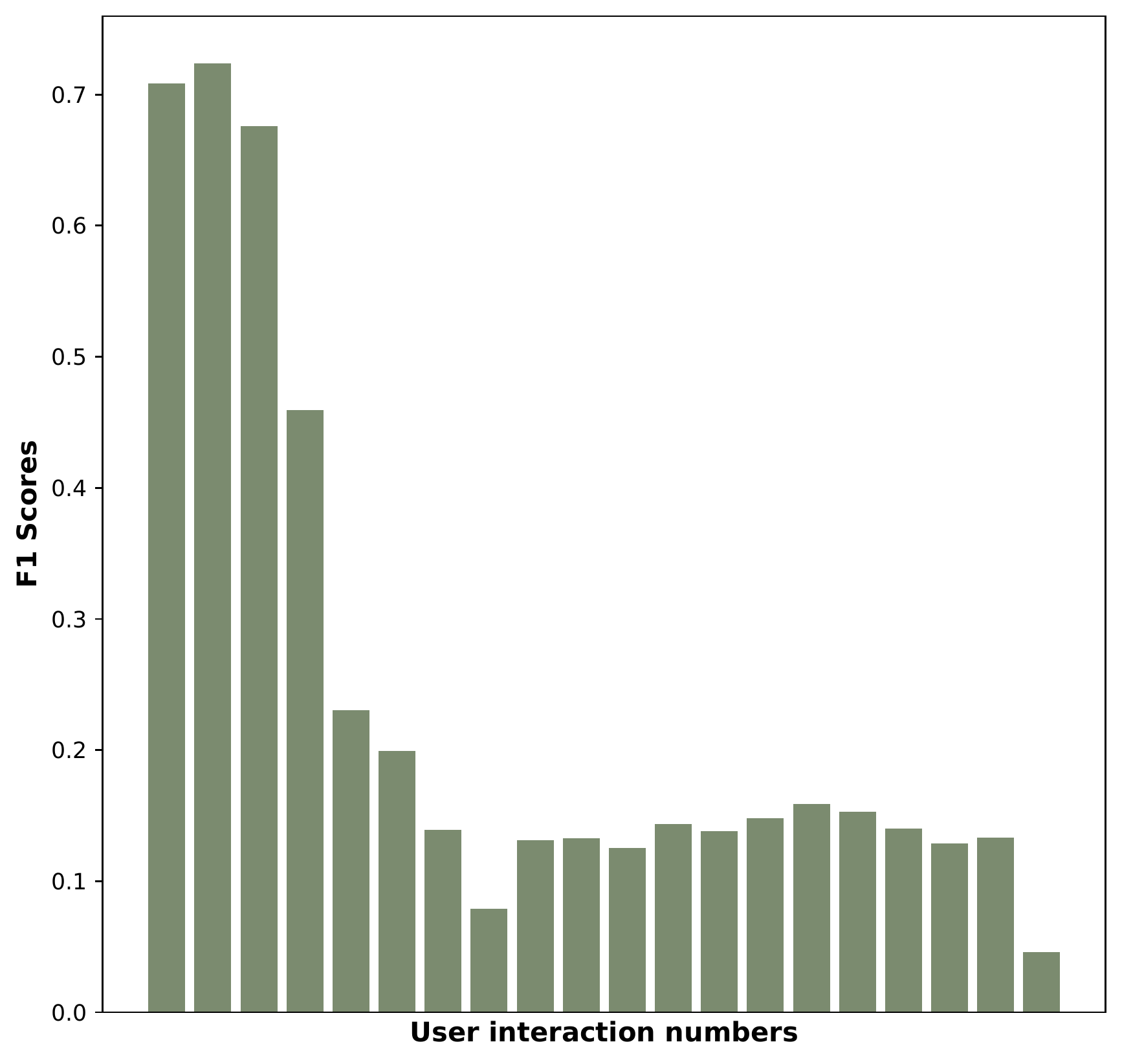}\label{fig_steam_lightgcn_interaction}}
  \caption{IMIA attacker performance for users with different number of interactions.}\label{fig_all_interaction}
\end{figure*}
\begin{figure}[ht]
  \centering
  \includegraphics[width=0.35\textwidth]{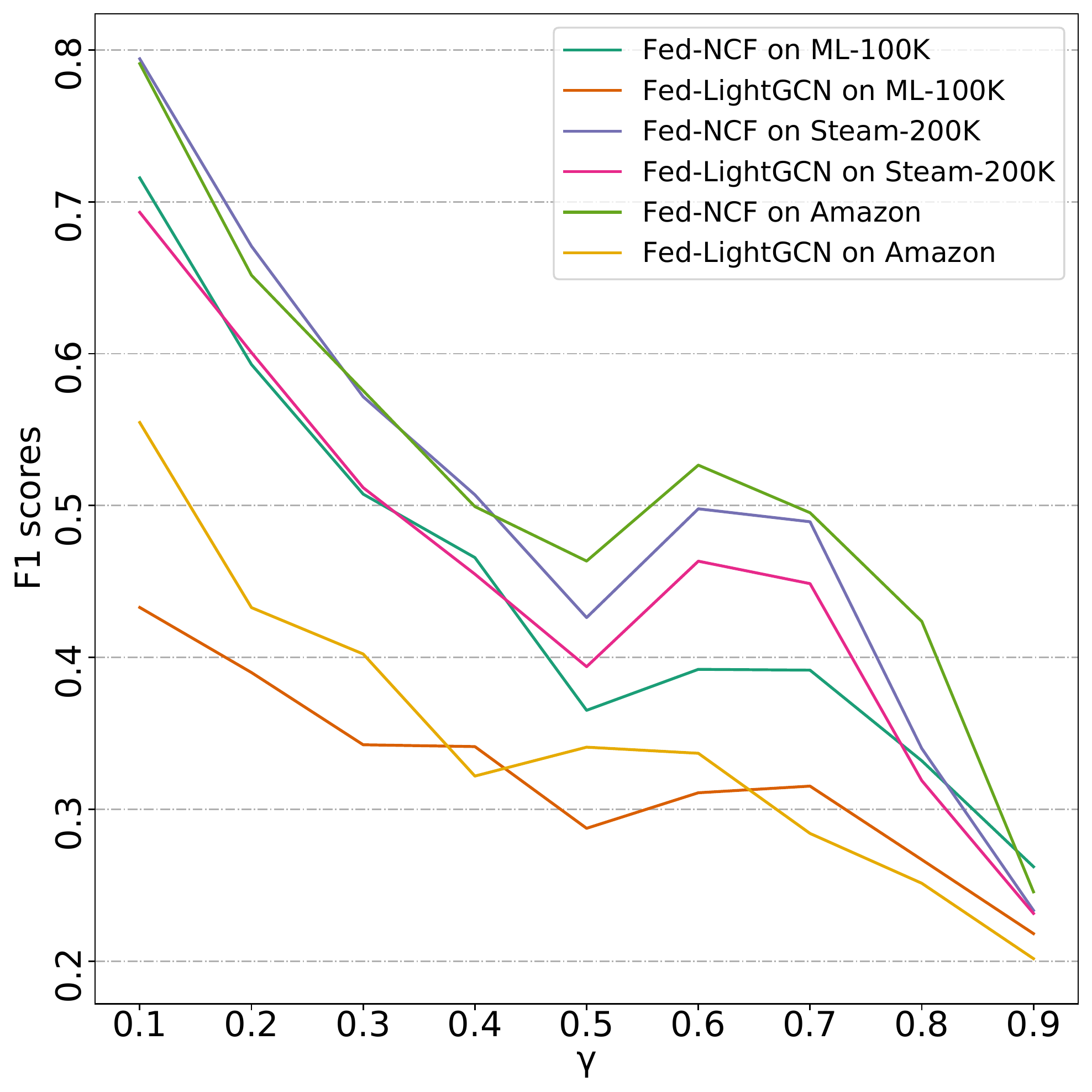}
  \caption{Our IMIA attacker's performance with different values of $\gamma$. $0.1$ means selecting the top $10\% * \left|\mathcal{V}_{i} \right|$ items as correct guesses according to distance metrics each iteration.}\label{fig_gamma}
\end{figure} 

\section{Details of Embeddings Deviation}\label{app_deviation}
In Section~\ref{sec_imia_def}, we present the trend of embeddings' deviation on MovieLens-100K.
Here, we calculate all FedRecs' embeddings deviation from their converged point to the initial point using L2 loss.
In Table~\ref{tb_deviation}, after applying our defense method, the deviation of item embedding is restrained, meanwhile, the user embedding is forced to update more.
As a result, more information is encoded in private parameters, rather than in public parameters.
Besides, across FedRecs, we can find that the updates of user embeddings are more significant in Fed-LightGCN than in Fed-NCF.
This observation is consistent with our argument that ``private parameters in Fed-LightGCN are more sufficiently used than in Fed-NCF''.

\section{The impact of interaction number}\label{app_interaction}
Fig.~\ref{fig_all_interaction} is an extension of Fig.~\ref{fig_user_interaction}.
We cluster users into $20$ groups based on their interaction numbers and report their average F1 score.
Since users in Amazon Cell Phone all have fewer interactions, we only visualize the statistics of MovieLens-100K and Steam-200K.
As shown in Fig.~\ref{fig_all_interaction}, users with fewer interactions are prone to leak more interaction information.
This phenomenon is more obvious in Fed-LightGCN, because by using convolution aggregation, users with more interaction will have more complicated private embeddings, therefore, they are difficult to be attacked by solely relying on public parameters.
This observation further implies that to prevent IMIA, we should improve the importance of private parameters.

\subsection{The Impact of $\gamma$}\label{app_gamma}
The hyper-parameter $\gamma$ denotes the percentage of items whose ratings the attacker is assumed to correctly infer at each iteration. 
Fig.~\ref{fig_gamma} illustrates the trend of the attacker's performance with different $\gamma$ on all datasets.
Generally, with smaller $\gamma$, the attacker achieves better performance.
For example, when $\gamma=0.1$, the attacker achieves nearly $0.8$ F1 scores on Fed-NCF and MovieLens-100K, however, when $\gamma=0.9$, the performance is reduced to lower than $0.3$.
On the other hand, smaller $\gamma$ needs more iterations to infer all the target user's interacted items. A desirable $\gamma$ value should make a good balance between attack effectiveness and attack efficiency.








\end{document}